\documentclass[conference]{IEEEtran}
\IEEEoverridecommandlockouts
\usepackage{cite}
\usepackage{graphicx}
\usepackage{xcolor}
\def\BibTeX{{\rm B\kern-.05em{\sc i\kern-.025em b}\kern-.08em
    T\kern-.1667em\lower.7ex\hbox{E}\kern-.125emX}}

\usepackage{mathtools,amsmath,amsfonts,amssymb}
\usepackage{algorithmic}
\usepackage{algorithm}
\usepackage{array}
\usepackage[caption=false,font=normalsize,labelfont=sf,textfont=sf]{subfig}
\usepackage{textcomp}
\usepackage{stfloats}
\usepackage{url}
\usepackage{verbatim}
\usepackage{graphicx}
\graphicspath{{Figures/}}
\usepackage{cite}
\usepackage{multirow}
\usepackage{acronym}
\usepackage{lipsum}
\usepackage{nccmath}

\begin{document}

\title{Enhanced Index-Based Feedback Overhead Reduction for WLANs
\thanks{This research is supported in part by the National Science Foundation (NSF) CNS through the award number 1814727.}}

\author{

\IEEEauthorblockN{Mrugen Deshmukh\IEEEauthorrefmark{1}\IEEEauthorrefmark{2}, Zinan Lin\IEEEauthorrefmark{2}, Hanqing Lou\IEEEauthorrefmark{2}, Mahmoud Kamel\IEEEauthorrefmark{3}, Rui Yang\IEEEauthorrefmark{2}, Ismail Guvenc\IEEEauthorrefmark{1}
}
\IEEEauthorblockA{\IEEEauthorrefmark{1}Department of Electrical and Computer Engineering, NC State University, Raleigh, NC 27606}
\IEEEauthorblockA{\IEEEauthorrefmark{2}InterDigital, Inc., New York, NY 10120}
\IEEEauthorblockA{\IEEEauthorrefmark{3}InterDigital, Inc., Montreal, QC H3A 3G4\\
Email:\{madeshmu,iguvenc\}@ncsu.edu, \{Zinan.Lin, Hanqing.Lou, Mahmoud.Kamel, Rui.Yang\}@InterDigital.com}
}

\maketitle

\begin{abstract}
Compressed beamforming algorithm is used in the current Wi-Fi standard to reduce the beamforming feedback overhead (BFO). However, with each new amendment of the standard the number of supported antennas in Wi-Fi devices increases, leading to increased BFO and hampering the throughput despite using compressed beamforming. In this paper, a novel index-based method is presented to reduce the BFO in Wi-Fi links. In particular, a \textit{k}-means clustering-based approach is presented to generate candidate beamforming feedback matrices, thereby reducing the BFO to only the index of the said candidate matrices. With extensive simulation results, we compare the newly proposed method with the IEEE 802.11be baseline and our previously published index-based method. We show approximately $54\%$ gain in throughput at high signal-to-noise (SNR) against the IEEE 802.11be baseline. Our comparison also shows approximately $4$ dB gain compared to our previously published method at the packet-error-rate (PER) of $0.01$ using MCS index $11$. Additionally, we also discuss the impact of the distance metric chosen for clustering as well as candidate selection on the link performance. 
\end{abstract}

\section{Introduction}

The continuous evolution of wireless technology is an imperative driven by an unrelenting demand for higher data rates. Each new iteration of a wireless technology includes new features to address these demands. Accordingly, the Wi-Fi standards have included new amendments to increase the throughput by supporting higher bandwidths, higher modulation orders, including spatial multiplexing and transmit beamforming (TxBF). 

Using TxBF is a vital approach to improve the SNR at the receiver in a link and enhance the link performance. However, for optimal use of TxBF accurate channel state information (CSI) is required, which is obtained using a channel sounding procedure. As Wi-Fi evolves and supports an increasing number of antennas, the beamforming feedback overhead (BFO) associated with acquiring the CSI increases. This rising overhead puts a significant strain on the throughput performance of a link. The current Wi-Fi standard may use the compressed beamforming feedback method \cite{stacey_book} to reduce the BFO. But with the expectation that future Wi-Fi generations will support a large number of antennas for higher throughput and lower latency, the existing BFO reduction scheme may not be sufficient to meet those goals. 

Overhead reduction has been a topic of interest in Wi-Fi since the introduction of TxBF in the standard. Several traditional works have addressed this issue by optimizing channel sounding parameters \cite{7442585, 8013252, nabetani2018novel}. In \cite{6666172}, reducing overhead for multi-user MIMO (MU-MIMO) systems is discussed considering implicit as well as explicit feedback methods, whereas \cite{7809634} focuses on implicit feedback methods. In recent literature, there has been a surge of interest in using ML-based tools to solve problems in the Wi-Fi domain \cite{9786784}, where works such as \cite{9448323} and \cite{9259366} have used neural network-based approaches to address the beamforming overhead reduction problem. 

This paper presents a novel index-based approach to reduce the BFO. Using an index-based approach, the BFO is compressed efficiently by reducing the size of feedback information to the index in a candidate set. In our previous work \cite{9860553}, we developed an index-based beamforming feedback method where the candidates are generated from clustering over a dataset of vectors containing indices of the angles used in compressed beamforming. In this work, however, we use a different representation of the data to generate the candidates. Specifically, we use the complex elements in the unitary steering matrices to generate our candidate set. Our simulation results show that using the complex elements directly captures the correlation between different data points while clustering and leads to an improvement in candidate generation,  leading to higher accuracy of the candidates and a significantly better error-rate performance than the one in \cite{9860553}. Additionally, we discuss the effect of choosing the distance metric on the clustering performance. And we also discuss how using different distance metrics for candidate generation and beamforming selection affects the link performance. 

The rest of this paper is organized as follows. In Section~\ref{sec:sys_model}, we describe our system model, after which we describe our index-based feedback overhead approach in Section~\ref{sec:index_bf}. In Section~\ref{sec:sim_results} our simulation results are presented and we conclude the paper in Section~\ref{sec:conclusion} and discuss future work.

\section{System Model} \label{sec:sys_model}

\subsection{Downlink Data Transmission}

We consider a Single User Multiple Input Multiple Output (SU-MIMO) downlink transmission between an access point (AP) and a Wi-Fi station (STA). In this downlink transmission, the AP is the beamformer that uses TxBF to transmit data to the STA, which is the beamformee. Mathematically, the received signal of a general MIMO system using TxBF can be given by \cite{stacey_book}
\begin{equation}\label{eq:downlink_Y}
    \mathbf{y} = \sqrt{\frac{\rho}{N_\text{r}}}\mathbf{H}\hat{\mathbf{V}}\mathbf{x} + \mathbf{z},
\end{equation}
where the received signal $\mathbf{y}$ is a column vector of size $N_\text{t}\times 1$, the transmitted signal $\mathbf{x}$ is a column vector of size $N_\text{c}\times 1$, $\mathbf{H}$ is the MIMO channel matrix of size $N_\text{t}\times N_\text{r}$ and the AWGN $\mathbf{z}$ is a vector of size $N_\text{t}\times 1$, and $\rho$ is the SNR. The dimensions of these matrices are as follows: $N_\text{r}$ is the number of antennas at the beamformer, $N_\text{t}$ is the number of antennas at the beamformee, and $N_\text{c}$ is the number of spatial streams being transmitted. Here, we assume $N_\text{c} = N_\text{t}$.

The beamforming or precoding matrix $\hat{\mathbf{V}}$ is of the size $N_\text{r}\times N_\text{c}$ and is generally obtained via a channel sounding procedure before the data transmission. In a typical Wi-Fi link, the channel-sounding operation entails the transmission of a Null Data Packet (NDP) frame from the beamformer to the beamformee. The beamformee will use the Long Training Field (LTF) frames within this NDP to estimate the channel response. From the Singular Value Decomposition (SVD) of this channel response, the beamformee will compute $\mathbf{V}$ and send the relevant information required for the beamformer to re-construct $\mathbf{V}$ in the beamforming report (BFR). The re-constructed $\mathbf{V}$ at the beamformer is labeled as $\hat{\mathbf{V}}$.

Traditional approaches used to transmit the beamforming feedback may be broadly divided into two parts - implicit beamforming and explicit beamforming \cite{11bedraft}. The explicit method used in 802.11be is compressed beamforming feedback. In this work, we present a novel index-based approach for beamforming feedback as an alternative to these traditional feedback methods, which will be discussed in Section~\ref{sec:index_bf}.

\subsection{Goodput Calculations}

The goodput of a Wi-Fi link may be considered as the effective error-free data packet transmission. The mathematical formula to calculate the goodput at a given SNR value may be represented as follows  
\begin{equation}\label{eq:goodput_2}
    \Gamma = \frac{L_{\text{data}}}{T_{\text{sounding}} + T_{\text{data}}/(1-P_\text{e}) + T_{\text{SIFS}} + T_{\text{ACK}}},
\end{equation}
where $P_\text{e}$ represents the Packet Error Rate (PER) at that particular SNR, $L_{\text{data}}$ is the length of the data payload, $T_{\text{sounding}}$ represents the channel sounding duration, $T_{\text{data}}$ is the data transmission duration, $T_{\text{SIFS}}$ is the short inter-frame spacing between completion of data transmission and reception of ACK signal at the beamformer, and $T_{\text{ACK}}$ is the duration required for ACK signal transmission. As can be noted in (\ref{eq:goodput_2}), high BFO may increase $T_{\text{sounding}}$ and subsequently have a detrimental effect on the goodput. 

For the calculations in (\ref{eq:goodput_2}), we assume that the sounding procedure is not repeated in cases where data packet re-transmission is required.

\section{Proposed Index-Based Beamforming Feedback} \label{sec:index_bf} 

In this section, we describe the proposed serialized-$\mathbf{V}$ index-based approach for beamforming feedback. The central idea behind index-based beamforming feedback is to reduce the BFO by replacing the traditional methods that transmit information about the $\hat{\mathbf{V}}$ matrix in the BFR with the index of a candidate within a candidate set. In the index-based feedback, after the beamformee computes the $\mathbf{V}$ from the SVD of the channel matrix estimate, it will compare $\mathbf{V}$ to the \textit{closest} candidate in the candidate set. In the BFR, the beamformee will transmit the index of the closest candidate instead of the information about the $\mathbf{V}$ matrix itself. In the proposed index-based beamforming feedback method, there are two crucial stages, namely, the candidate set generation and the beamforming feedback generation (selection of the closest candidate), which we describe in the following subsections. 

\begin{figure}[t] 
    \centering\vspace{-3mm}
    \includegraphics[width=0.99\linewidth,trim={8cm 6cm 8cm 6cm},clip]{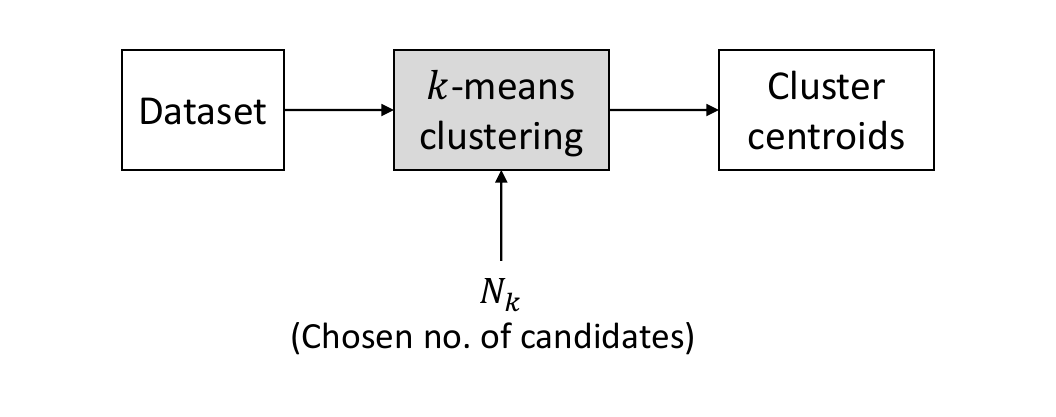}\vspace{-4mm}
    \caption{Candidate set generation procedure}
    \label{fig:candidate_generation}\vspace{-4mm}
\end{figure}

\subsection{Candidate Generation}

The candidate generation in our proposed method is done using an \textit{offline}-learning approach. The general flow of the procedure for candidate generation is shown in Fig.~\ref{fig:candidate_generation}. We generate a large dataset that contains data points from a diverse set of channel conditions. Consider a $N_{\text{r}}\times N_{\text{c}}$ MIMO case, where the $\mathbf{V}$ matrix after SVD of the channel matrix can be represented as 
\begin{equation}\label{eq:v_mat_form}
    \mathbf{V} = \begin{bmatrix} 
         v_{1,1} & v_{2,1} & \hdots & v_{N_{\text{r}},1} \\ 
         \vdots & \hdots & \hdots & \vdots \\ 
         v_{1,N_{\text{c}}} & v_{2,N_{\text{c}}} & \hdots & v_{N_{\text{r}},N_{\text{c}}}  
        \end{bmatrix}^T,
\end{equation}
where each $v_{i,j}$ represents a complex element of the $\mathbf{V}$ matrix for the $i$th row and $j$th column respectively. Next, we serialize this $\mathbf{V}$ matrix so that we have a single vector instead of a matrix. This vectorized form of $\mathbf{V}$ may be expressed as 
\begin{equation}
\label{eq:serialV}
    \mathbf{v}_{\text{s}} = [v_{1,1}, v_{2,1}, \hdots, v_{N_{\text{r}},1}, \hdots, v_{1,N_{\text{c}}}, v_{2,N_{\text{c}}}, \hdots, v_{N_{\text{r}},N_{\text{c}}}]^T.
\end{equation}

In our dataset, each data point is represented by such a vector. After the dataset is collected, it is fed to the \textit{k}-means clustering algorithm \cite{kmeans_arthur}. The \textit{k}-means clustering algorithm iterates through the dataset and groups it into $N_{\text{k}}$ clusters, where $N_{\text{k}}$ is a user-defined number. After the \textit{k}-means algorithm converges, we re-arrange the cluster centroids from the vector form to the matrix form as in (\ref{eq:v_mat_form}) and make the columns orthogonal using the Gram-Schmidt procedure \cite{strang_algebra}. These matrices are then used as our candidate set. The candidate set thus generated is assumed to be available at both the beamformer and the beamformee.

The distance metric used while clustering plays a crucial role in the generation of candidates. In our method, we consider two distance metrics - 1) Squared Euclidean distance (SED) and 2) Cosine distance (CD). Consider two serialized vectors $\mathbf{v}_{\text{s,a}}$ and $\mathbf{v}_{\text{s,b}}$. The CD between these two vectors can be calculated as 
\begin{align}
    d_{\text{CD}} &= 1 - \frac{\lvert \mathbf{v}_{\text{s,a}}^H\mathbf{v}_{\text{s,b}} \rvert}{\sqrt{(\mathbf{v}_{\text{s,a}}^H\mathbf{v}_{\text{s,a}})(\mathbf{v}_{\text{s,b}}^H\mathbf{v}_{\text{s,b}})}} \\
	&= 1 - \frac{\lvert \mathbf{v}_{\text{s,a}}^H\mathbf{v}_{\text{s,b}} \rvert}{N_{\text{c}}^2},
\label{eq:sv_cd}
\end{align}
where $\mathbf{v}_{\text{s,a}}$ and $\mathbf{v}_{\text{s,b}}$ are both serialized forms of unitary matrices, therefore $\mathbf{v}_{\text{s,a}}^H\mathbf{v}_{\text{s,a}} = \mathbf{v}_{\text{s,b}}^H\mathbf{v}_{\text{s,b}} = N_{\text{c}}$, where $N_{\text{c}}$ is the number of columns in the corresponding $\mathbf{V}$ matrix. 

The SED between these two vectors $\mathbf{v}_{\text{s,a}}$ and $\mathbf{v}_{\text{s,b}}$, on the other hand, can be calculated as 
\begin{align}
    d_{\text{SED}} &=  \lvert \mathbf{v}_{\text{s,a}} - \mathbf{v}_{\text{s,b}} \rvert ^2 \label{eq:sv_se_0}\\
	&= (\mathbf{v}_{\text{s,a}} - \mathbf{v}_{\text{s,b}})^H(\mathbf{v}_{\text{s,a}} - \mathbf{v}_{\text{s,b}}) \nonumber\\
	&= \mathbf{v}_{\text{s,a}}^H\mathbf{v}_{\text{s,a}} + \mathbf{v}_{\text{s,b}}^H\mathbf{v}_{\text{s,b}} - \mathbf{v}_{\text{s,a}}^H\mathbf{v}_{\text{s,b}} - \mathbf{v}_{\text{s,b}}^H\mathbf{v}_{\text{s,a}} \nonumber\\
      &= N_{\text{c}} + N_{\text{c}} - (\mathbf{v}_{\text{s,a}}^H\mathbf{v}_{\text{s,b}} + (\mathbf{v}_{\text{s,a}}^H\mathbf{v}_{\text{s,b}})^H) \nonumber \\
	&= 2\times N_{\text{c}} - 2\times \text{real}(\mathbf{v}_{\text{s,a}}^H\mathbf{v}_{\text{s,b}}). \label{eq:sv_sed}
\end{align}

Comparing (\ref{eq:sv_cd}) and (\ref{eq:sv_sed}), it can be seen that the CD between $\mathbf{v}_{\text{s,a}}$ and $\mathbf{v}_{\text{s,b}}$ is inversely proportional to the inner product of the two vectors, whereas the SED between $\mathbf{v}_{\text{s,a}}$ and $\mathbf{v}_{\text{s,b}}$ is inversely proportional to the real part of the inner product of the two vectors. Hence, we note that using CD we can capture the pair-wise correlation between the two vectors more accurately than using the SED. Thus, using CD while clustering (candidate generation) and while candidate selection (beamforming feedback generation) will lead to better link performance, as will be demonstrated in Section~\ref{sec:sim_results}.

\subsection{Beamforming Feedback Generation}

The beamforming feedback generation in the index-based method is performed during the channel sounding procedure at the beamformee using the following steps - 

\begin{enumerate}
	\item The beamformee computes the $\mathbf{V}$ from the SVD of the channel matrix estimate.
	\item The beamformee determines $\mathbf{v}_{\text{s}}$ the serialized version of $\mathbf{V}$ as in (\ref{eq:serialV}).
	\item The candidates are represented in the serialized form, and the beamformee finds the closest candidate to $\mathbf{v}_{\text{s}}$ using $d_{\text{CD}}$ or $d_{\text{SED}}$.
	\item In the BFR, the beamformee reports the index of the closest candidate. 
	\item After receiving the BFR, the beamformer will use the index to determine the candidate beamforming matrix, i.e., $\hat{\mathbf{V}}$ in (\ref{eq:downlink_Y}). 
	\item The beamformer may use this candidate for TxBF.
\end{enumerate}

In step 3 above, the candidates are converted to serial form before finding the closest candidate to $\mathbf{v}_{\text{s}}$ in order to be consistent with how the \textit{k}-means clustering algorithm operates for candidate generation.

\section{Simulation Results} \label{sec:sim_results} 


\subsection{Simulation Setup}

In our link-level simulation, an SU-MIMO downlink transmission is considered. To compare the performance of the proposed method we use the following two baselines: IEEE 802.11be standard using compressed beamforming, and our previously proposed index-based feedback approach iFOR \cite{9860553}. For compressed beamforming, we use $6$ bits to quantize the $\phi$ angles and $4$ bits to quantize the $\psi$ angles \cite{11bedraft}. We use the toolboxes offered by MathWorks in all our simulations.

For generating our dataset, we use channel models A-E defined by the IEEE 802.11 working group \cite{chan_model}. We use combined data generated from these channel models to generate our candidate sets using \textit{k}-means clustering. The link-level simulation results (e.g. PER), however, are obtained using the channel model D only, unless mentioned otherwise. We consider $N_\text{r} = 8$ and $N_\text{c} = 2$ (each $\hat{\mathbf{V}}$ is of the size $8\times 2$), and a channel bandwidth of $20$ MHz, which contains $242$ subcarriers. For all index-based feedback approaches, $10$ bits of feedback per subcarrier group is used, which represents $1024$ candidates. The payload in each packet is $1,000$ bytes. All the PER and intermediate KPI results are obtained after averaging over $10,000$ packets. We consider subcarrier grouping $N_\text{g} = 4$ for reporting the beamforming feedback. In the legend of the figures in this section, FB represents feedback.

\begin{table}[t]
  \centering
  \caption{Simulation results for intermediate KPIs}
  \label{tab:kpi_results}
  \begin{tabular}{|p{0.25\linewidth}|p{0.2\linewidth}|p{0.15\linewidth}|p{0.15\linewidth}|}
    \hline
    \textbf{Method} & \textbf{KPI} & \textbf{Chan. B} & \textbf{Chan. D} \\
    \hline
    \multirow{2}{*}{Serialized-$\mathbf{V}$ (CD)}    & $\rho$ & 0.8288 & 0.7938 \\
    \cline{2-4} 			     					  	& NMSE (dB) & 1.3212 & 1.6207 \\
    \hline
     \multirow{2}{*}{Serialized-$\mathbf{V}$ (SED)} & $\rho$ & 0.8049 & 0.7705 \\
    \cline{2-4} 			     						& NMSE (dB) & -2.5741 & -1.9221 \\ 
    \hline  
  \end{tabular}
\end{table}
\subsection{Intermediate KPIs}

Intermediate KPIs such as the Normalized Mean Square Error (NMSE) or the Generalized Cosine Similarity ($\rho$) allow for quick evaluation of the accuracy of the feedback generated using a candidate set compared to the actual feedback matrix. 

\subsubsection{Normalized Mean Square Error (NMSE)}

Mathematically, the NMSE can be calculated as 
\begin{equation}\label{eq:nmse}
	\text{NMSE} = \mathbb{E}\Bigg\{ \frac{ \lVert \mathbf{v}_{\text{s}} - \mathbf{v}_{\text{c}} \rVert^2_2}{ \lVert \mathbf{v}_{\text{s}} \rVert^2_2} \Bigg\},
\end{equation}
where $\mathbf{v}_{\text{s}}$ is the serialized form of $\mathbf{V}$, $\mathbf{v}_{\text{c}}$ represents the chosen candidate of which the index is sent back in the BFR in serialized form, and $\lVert . \rVert_2$ represents the $L_2$-norm of the corresponding entry.

\subsubsection{Generalized Cosine Similarity ($\rho$)}

To determine the similarity between $\mathbf{V}$ and the chosen candidate matrix $\mathbf{V}_{\text{c}}$, $\rho$ can be calculated as 
\begin{equation}\label{eq:gcs}
	\rho = \mathbb{E}\Bigg\{\frac{1}{N_{\text{SC}}}\sum_{n=1}^{N_{\text{SC}}} \frac{\lvert \mathbf{v}_{\text{c},1}^H {\mathbf{v}}_1 \rvert}{\lVert \mathbf{v}_{\text{c},1} \rVert^2_2\lVert {\mathbf{v}}_1 \rVert^2_2} \Bigg\}
\end{equation}
where ${\mathbf{v}}_1$ and $\mathbf{v}_{\text{c},1}$ represent the first columns of ${\mathbf{V}}$ and $\mathbf{V}_{\text{c}}$ respectively, and $N_{\text{SC}}$ is the number of subcarriers. Adhering to the methodology in \cite{8322184}, we calculate the $\rho$ using only the first columns of ${\mathbf{V}}$ and $\mathbf{V}_{\text{c}}$.

Table~\ref{tab:kpi_results} shows the KPI results for the serialized-$\mathbf{V}$ method using CD and SED. To obtain these results, we simulate the channel sounding procedure without additive noise, perform SVD on the channel estimate $\mathbf{H}$ to obtain the ${\mathbf{V}}$ matrix, and then compare the accuracy of the closest candidate using either of the serialized-$\mathbf{V}$ methods with this ${\mathbf{V}}$. The simulations are performed using channel models B and D.

\begin{figure}[t] 
    \centering\vspace{-3mm}
    \includegraphics[width=0.99\linewidth]{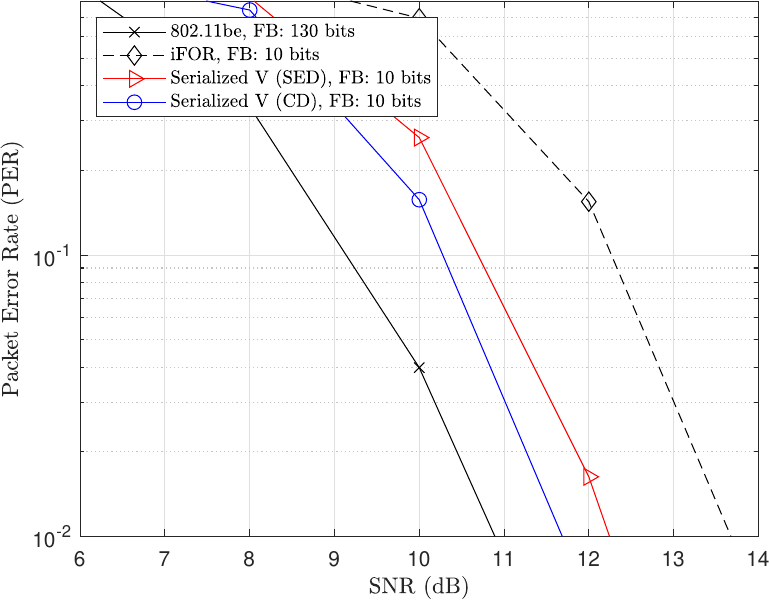}\vspace{-1mm}
    \caption{PER versus SNR comparison for MCS index 3}
    \label{fig:per_mcs3}\vspace{-4mm}
\end{figure}

Comparing (\ref{eq:sv_cd}) and (\ref{eq:gcs}), it can be said that the GCS ($\rho$) is analogous to CD given that they are both proportional to the inner product of the respective vectors. Similarly, comparing (\ref{eq:sv_se_0}) and (\ref{eq:nmse}), the NMSE is analogous to SED given that they are both directly proportional to the square of the difference between the respective vectors. Correspondingly, it is clear from Table~\ref{tab:kpi_results} that using the serialized-$\mathbf{V}$ with CD yields higher $\rho$ than when SED is used, and the NMSE is lower when the  serialized-$\mathbf{V}$ with SED is used. These results give an indication of what to expect in the PER performance, as we will discuss in the later subsections.

\begin{figure}[ht] 
    \centering\vspace{-3mm}
    \includegraphics[width=0.99\linewidth]{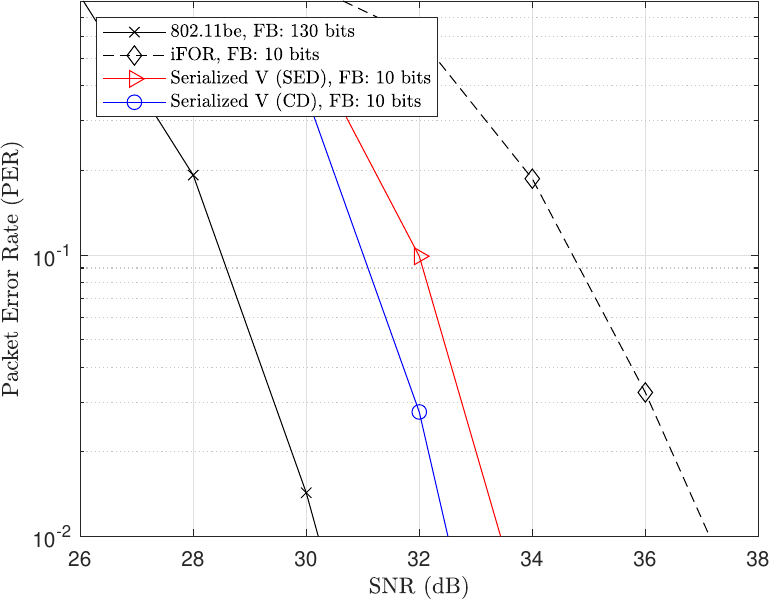}\vspace{-1mm}
    \caption{PER versus SNR comparison for MCS index 11}
    \label{fig:per_mcs11}\vspace{-4mm}
\end{figure}
It can also be observed that the channel condition is an important factor in the quality of the KPIs. The simulations in channel model D incur more severe multipath fading compared to channel model B. Hence, the results for channel model D show lower $\rho$ and higher NMSE for both approaches. Another important factor for our proposed method is the size of the candidate set being used. Using a candidate set-based approach results in a loss in accuracy of the $\hat{\mathbf{V}}$ matrix. A higher size of the candidate set may lower the loss in accuracy leading to better link performance, including better KPI results at the cost of higher complexity. Performance evaluation for different sizes of candidate sets is out of the scope of this paper. 
\subsection{PER Comparison} \label{sec:per_comparison}

We compare the PER performance for MCS index 3 and 11 in Fig.~\ref{fig:per_mcs3} and Fig.~\ref{fig:per_mcs11} respectively. MCS index 3 represents $16$-QAM and LDPC code with code rate $1/2$, whereas MCS index $11$ represents $1024$-QAM and a code rate of $5/6$. As can be seen from both the figures, using the serialized-$\mathbf{V}$ approach with both CD and SED yields better performance than that of iFOR. In iFOR, the candidates are generated by clustering over the vectors of the indexes of quantized angles representing the $\mathbf{V}$ matrix using SED as the distance metric. In the serialized-$\mathbf{V}$ approach, however, we cluster over vectors of the complex elements in $\mathbf{V}$, capturing the physical meaning of the $\mathbf{V}$ matrix in a more meaningful way leading to diverse and efficient candidate set. 

In the serialized-$\mathbf{V}$ approach, using CD as the distance metric is more appropriate to capture the pair-wise correlation between two different vectors while clustering as discussed in (\ref{eq:sv_cd}) and (\ref{eq:sv_sed}) in Section~\ref{sec:index_bf}. Using the candidates obtained with CD as the distance metric thus leads to better PER performance. For MCS index $3$, the degradation in PER using serialized-$\mathbf{V}$ with CD compared to 802.11be is less than $1$ dB despite only using $10$ bits per subcarrier group against the $130$ per subcarrier group in 802.11be. For MCS index $11$, the degradation in PER using serialized-$\mathbf{V}$ with CD compared to 802.11be is approximately $2$ dB, whereas it has a gain of approximately $4$ dB compared to the iFOR method. 

Thus, using the vectorized representation of the complex elements of $\mathbf{V}$, we are able to achieve significantly higher PER performance compared to our previously proposed method for the same amount of information in the beamforming feedback.

\subsection{MCS Selection}\label{sec:mcs_selection}

\begin{figure}[t] 
    \centering\vspace{-3mm}
    \includegraphics[width=0.99\linewidth]{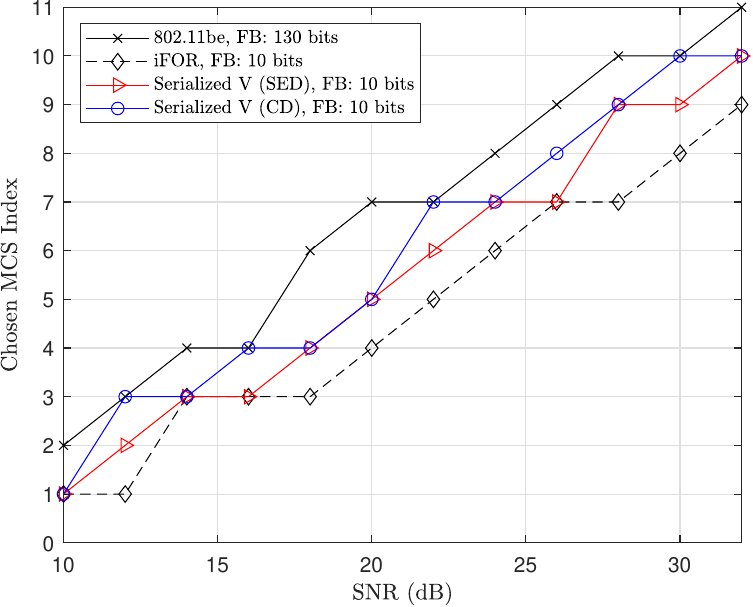}\vspace{-1mm}
    \caption{MCS selection versus SNR for $8\times2$ MIMO}
    \label{fig:mcs_selection}\vspace{-4mm}
\end{figure}

In Fig.~\ref{fig:mcs_selection}, we show the chosen MCS index for a given SNR in our simulations. To obtain these results, we perform PER simulations for MCS indexes $0-11$. Then, for a given SNR we choose the highest MCS index that has $P_e \leq 0.01$. A particular MCS index corresponds to the modulation and the code rate (LDPC code in our simulations), as listed in Table~\ref{tab:mcs_table}. Hence, the chosen MCS index is an important factor that determines the data rate being chosen for transmitting data, with a higher chosen MCS index leading to a higher data rate. 

As can be noted in Fig.~\ref{fig:mcs_selection}, using the IEEE 802.11be baseline generally leads to the highest chosen MCS index across all SNR points, given that it has the best PER performance. The iFOR method, on the other hand, has the worst PER performance and generally chooses the lowest MCS index. Using either of the serialized-$\mathbf{V}$ methods leads to better PER performance than iFOR and worse PER performance than the 802.11be baseline, which is also corroborated in the MCS index selection. For the serialized-$\mathbf{V}$ methods, using CD leads to better PER performance than when SED is used, which is why using CD generally a higher MCS index is chosen. 

\begin{table}[b]
  \centering
  \caption{Modulation and code rate for a given MCS index}
  \label{tab:mcs_table}
  \begin{tabular}{|p{0.075\linewidth}|p{0.15\linewidth}|p{0.1\linewidth}|p{0.075\linewidth}|p{0.15\linewidth}|p{0.1\linewidth}|}
    \hline
    \textbf{Index} & \textbf{Modulation} & \textbf{Code Rate} & \textbf{Index} & \textbf{Modulation} & \textbf{Code Rate} \\
    \hline  
    0 & BPSK & 1/2 & 6 & 64-QAM & 3/4 \\
    \hline  
    1 & QPSK & 1/2 & 7 & 64-QAM & 5/6 \\
    \hline 
    2 & QPSK & 3/4 & 8 & 256-QAM & 3/4 \\
    \hline 
    3 & 16-QAM & 1/2 & 9 & 256-QAM & 5/6 \\
    \hline 
    4 & 16-QAM & 3/4 & 10 & 1024-QAM & 3/4 \\
    \hline 
    5 & 64-QAM & 2/3 & 11 & 1024-QAM & 5/6 \\
    \hline 
  \end{tabular}
\end{table}
\subsection{Goodput Comparison}

Looking back to the goodput calculations in (\ref{eq:goodput_2}), three variables play a role in determining the goodput performance - 1) $P_e$: the PER simulated at a given SNR value, 2) $T_{\text{data}}$: the time required for data transmission and 3) $T_{\text{sounding}}$: the time required for channel sounding.

Now, $P_e$ itself does not play a huge role in the goodput in our set of simulations, as the target $P_e$ is $0.01$. But it affects the MCS selection, and the chosen MCS index impacts the chosen data rate which determines $T_{\text{data}}$. Hence, a higher chosen MCS index leads to a reduction in $T_{\text{data}}$ and leads to an improvement in the goodput. 

\begin{figure}[t] 
    \centering\vspace{-3mm}
    \includegraphics[width=0.99\linewidth]{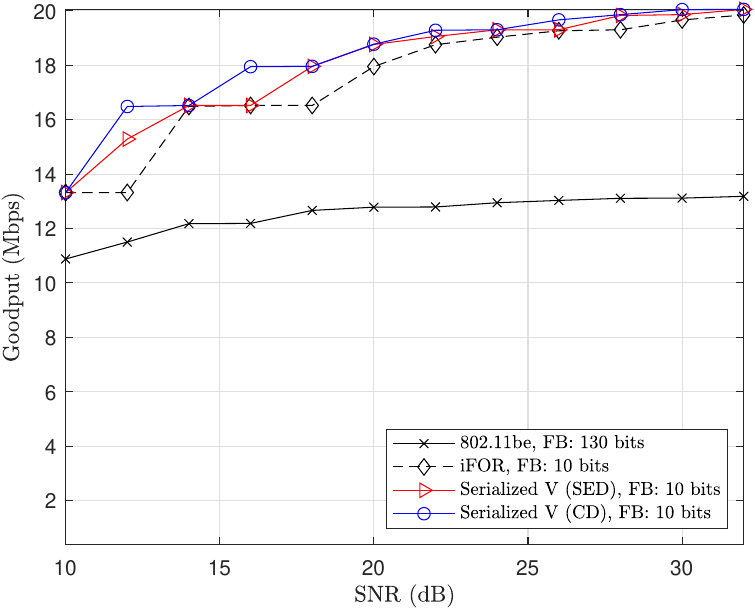}\vspace{-1mm}
    \caption{Goodput comparison versus SNR for a payload of $1,000$ bytes}
    \label{fig:gput_comparison}\vspace{-4mm}
\end{figure}

The most crucial variable in our goodput calculations is the channel sounding duration ($T_{\text{sounding}}$). The motivation in the index-based beamforming feedback is to reduce $T_{\text{sounding}}$ so that the goodput increases significantly. In our simulations, we use $10$ bits of feedback per subcarrier group for the index-based methods instead of the $130$ bits per subcarrier group used in IEEE 802.11be. This is why, looking at the goodput results in Fig.~\ref{fig:gput_comparison}, it is clear that despite the loss in $P_e$ and lower chosen MCS index, the reduction in $T_{\text{sounding}}$ using the index-based feedback methods results in high goodput gain, with the highest gain being approximately $54\%$ at the SNR of $32$ dB using the serialized-$\mathbf{V}$ method with CD.

In Section~\ref{sec:per_comparison}, we show that using the serialized-$\mathbf{V}$ approach we can achieve significant gain in the PER performance compared to iFOR. But in the goodput calculations, since the PER gain is not a significant factor, the goodput performance using the iFOR method is comparable to the proposed serialized-$\mathbf{V}$ approach despite worse PER performance and lower MCS index selection. The iFOR method also uses $10$ bits of feedback per subcarrier group. Even for the  serialized-$\mathbf{V}$ approach using SED, the goodput performance is comparable to the serialized-$\mathbf{V}$ approach using CD despite noticeable differences in the PER performance. 

\subsection{Effect of Distance Metric on Link Performance}

\begin{figure}[t] 
    \centering\vspace{-3mm}
    \includegraphics[width=0.99\linewidth]{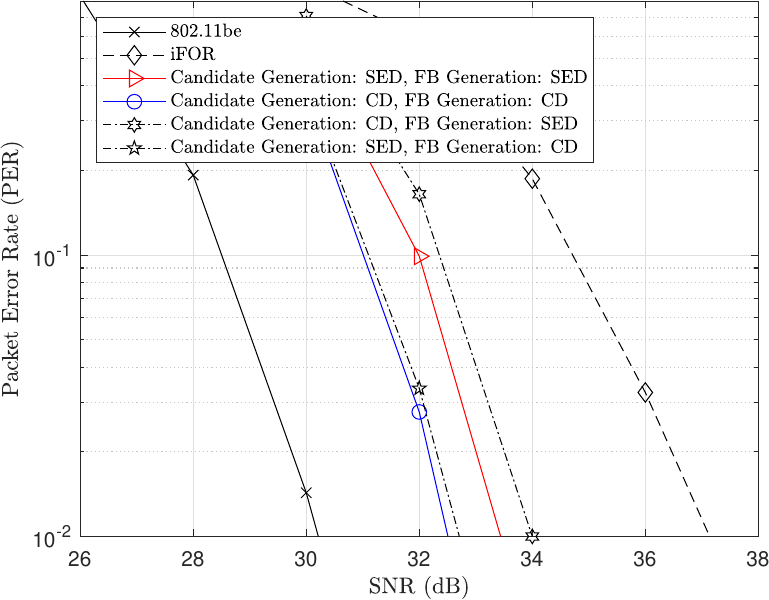}\vspace{-1mm}
    \caption{Effect of choosing different distance metrics on PER}
    \label{fig:per_cross}\vspace{-4mm}
\end{figure}

In the serialized-$\mathbf{V}$ methods we have discussed so far, we consider that the same distance metric (either SED or CD) is used for both stages: 1) candidate generation and 2) beamforming feedback generation (discussed in Section~\ref{sec:index_bf}). However, depending on the implementation, using different distance metrics for the two different phases may have implications for the link performance. In Fig.~\ref{fig:per_cross}, we discuss the effects of changing the distance metric for the two phases on the PER using the MCS index $11$. We present results for the combination of the two distance metrics with the two different stages, resulting in four PER results.

Observing the PER performance of the four different curves in Fig.~\ref{fig:per_cross}, it can be noted that using SED for FB generation leads to sub-optimal PER performance. This provides further evidence for the observations made from (\ref{eq:sv_cd}) and (\ref{eq:sv_sed}) in Section~\ref{sec:index_bf}. The inconsistency of using CD for candidate generation and SED for feedback generation leads to further degradation in the PER. Using CD for FB generation, however, leads to the best PER performance, irrespective of the distance metric used for \textit{k}-means clustering while candidate generation. The results here indicate that using both CD and SED as distance metrics for candidate generation using \textit{k}-means clustering may lead to sufficient diversity in the candidate vectors but for beamforming feedback generation, the closest candidate must be selected using CD. Another notable observation in Fig.~\ref{fig:per_cross} is that all four results are better than that of our previously proposed iFOR method.

\section{Conclusion and Future Work} \label{sec:conclusion}

In this paper, we propose a novel index-based approach to reduce the beamforming feedback overhead in Wi-Fi systems. To generate the candidate set used in this approach, we vectorize the beamforming matrices and store them in a large dataset. This dataset is then fed to the \textit{k}-means clustering algorithm. After the clustering converges, we use the cluster centroids as our candidates. 

Our simulation results show that, if we use this approach instead of using candidates in the form of vectors of angle indices as is done in our previous work iFOR \cite{9860553}, we are able to capture the pair-wise correlation between two vectors more accurately and we can achieve a significant gain in the error-rate performance. Specifically, we show a gain in PER over iFOR of approximately $4$ dB when MCS index $11$ is used, and approximately $2$ dB when MCS index $3$ is used (at PER of $0.01$). Furthermore, even though there is a significant gain in the PER compared to our previous work, we observe that the improvement over iFOR in goodput performance is noticeable although not significant. This is because the reduction in beamforming overhead is similar in both index-based approaches, and it plays a crucial role in goodput calculations. 

Finally, we observe that using CD for candidate selection is an important factor in determining the PER performance, irrespective of whether CD or SED is used for candidate generation. For our future work, we plan to explore more ways to represent the data in our candidate sets and subsequently, their effect on the link performance.


\bibliographystyle{IEEEtran}

\bibliography{ref}


\end{document}